\documentclass[pdflatex,sn-mathphys]{sn-jnl}

\jyear{2021}%

\theoremstyle{thmstyleone}%
\theoremstyle{thmstyletwo}%

\theoremstyle{thmstylethree}%

\begin{document}

\title[Terahertz Radiation]{Intensity-surged and Bandwidth-extended Terahertz Radiation in Two-foci Cascading Plasmas}


\author[1,2]{\fnm{Yizhu} \sur{Zhang}}
\equalcont{These authors contributed equally to this work.}

\author[3]{\fnm{Zhi-Hong} \sur{Jiao}}
\equalcont{These authors contributed equally to this work.}

\author[2]{\fnm{Tao} \sur{He}}
\equalcont{These authors contributed equally to this work.}

\author[2,6]{\fnm{Jingjing} \sur{Zhao}}

\author[1]{\fnm{Xingwang} \sur{Fan}}

\author[2,6]{\fnm{Taotao} \sur{Chen}}

\author[3]{\fnm{Guo-Li} \sur{Wang}}
\author[2]{\fnm{Tian-Min} \sur{Yan}}
\author*[5]{\fnm{Xiao-Xin} \sur{Zhou}}\email{zhouxx@nwnu.edu.cn}

\author*[4,2,6]{\fnm{Yuhai} \sur{Jiang}}\email{jiangyh3@shanghaitech.edu.cn}

\affil[1]{\orgdiv{Center for Terahertz Waves and School of Precision Instrument and Optoelectronics Engineering}, \orgname{Tianjin University}, \orgaddress{
\postcode{300072},
\state{Tianjin}, \country{China}}}

\affil[2]{\orgdiv{Shanghai Advanced Research Institute}, \orgname{Chinese Academy of Sciences}, \orgaddress{
\postcode{201210}, \state{Shanghai}, \country{China}}}

\affil[3]{\orgdiv{Key Laboratory of Atomic and Molecular Physics and Functional Materials of Gansu Province, College of Physics and Electronic
Engineering}, \orgname{Northwest Normal University}, \orgaddress{
\postcode{730070}, \state{Lanzhou}, \country{China}}}
\affil[4]{
\orgname{Center for Transformative Science and School of Physical Science and Technology, ShanghaiTech University}, \orgaddress{
\postcode{201210}, \state{Shanghai}, \country{China}}}
\affil[5]{
\orgdiv{Beijing National Laboratory for Condensed Matter Physics},
\orgname{Institute of Physics, Chinese Academy of Sciences}, \orgaddress{
\postcode{100190}, \state{Beijing}, \country{China}}}
\affil[6]{
\orgname{University of Chinese Academy of Sciences}, \orgaddress{
\postcode{100049}, \state{Beijing}, \country{China}}}


\abstract{The two-color strong-field mixing in gas medium is a widely-used approach to generate bright broadband terahertz (THz) radiation. Here, we present a new and counterintuitive method to promote THz performance in two-color scheme. Beyond our knowledge that the maximum THz generation occurs with two-color foci overlapped, we found that, when the foci of two-color beams are noticeably separated along the propagation axis resulting in cascading plasmas, the THz conversion efficiency is surged by one order of magnitude and the bandwidth is stretched by more than 2 times, achieving $10^{-3}$ conversion efficiency and $>$100 THz bandwidth under the condition of 800/400 nm, $\sim$35 fs driving lasers. 
With the help of the pulse propagation equation and photocurrent model, the observations can be partially understood by the compromise between THz generation and absorption due to the spatial redistribution of laser energy in cascading plasmas. Present method can be extended to mid-infrared driving laser, and the new records of THz peak power and conversion efficiency are expected. }




\maketitle

\section{Introduction}\label{sec1}

The temporal and spatial mixing of strong fundamental frequency ($\omega$) and second harmonic ($2\omega$) laser fields in gas-phase medium can produce a ultrashort terahertz (THz) pulse \cite{Cook00}. Given the absence of emitter damage and absorption, the two-color scheme is particularly promising for intense and broadband THz radiation \cite{Koulouklidis2020, Mitrofanov2020,Jang2019}. The wonderful characteristics of the THz source has already been used in time-resolved terahertz spectroscopy \cite{Pashkin2011,Valverde-Chavez2015,Wang2016} and transient absorption spectroscopy \cite{Chen2016}. The bright broadband THz radiation has potential applications in broadband spectroscopy \cite{Cossel17}, atomic and molecular ultrafast imaging \cite{Zhang2018a}, advanced accelerator technique \cite{Zhang2018b}, etc.

Although the THz generation in two-color scheme has already been a widely-used, well-established technique, yet the attempts to improve the technique are never abandoned. In experimental setups, multiple control knobs have been tuned to increase the THz conversion efficiency and bandwidth. Firstly, the appropriate parameters of driving lasers, including the wavelength and pulse duration, can optimize the THz generation. The wavelength scaling investigations \cite{Clerici2013,Nguyen2019} exhibited that the driving laser with longer wavelength strongly enhances the THz down-conversion efficiency by one or two orders of magnitude comparing to THz generation driven by 800 nm pulses at same input energies. In practical experiments, mid-infrared fields at 3.9 $\mu$m delivered by optical parametric amplifiers have been used to produce extremely strong THz field above 100 MV/cm \cite{Koulouklidis2020, Mitrofanov2020,Jang2019}. Simultaneously, tuning the wavelength ratio of two-color fields \cite{Vvedenskii2014,Kostin2016,Zhang2017} can manipulate the central wavelength and effectively extend the THz bandwidth \cite{Thomson2010,Babushkin2011,Balciunas2015}. If more than two-color fields are involved in the process, multiple wavelength fields can also improve the THz performance \cite{Martinez2015}. 
Short pulses by means of pulse compression technique significantly broadens the bandwidth of THz pulse, yielding  supercontinuum radiation of bandwidth $>$100 THz by using 7 fs driving pulse \cite{Thomson2010,Matsubara2012,Blank2013}.
In addition, controlling the polarization of two-color fields is also necessary. It has been well acknowledged that the relative polarization of two-color lasers should be optimized in THz conversion process \cite{Xie2006,Zhang2020}. And two-color co-rotating circularly-polarized fields can further amplify THz emission by a factor of 5 comparing to linearly polarized fields \cite{Meng2016}. 
Finally, the plasma profile also plays a role on THz generation. By manipulating the focal length \cite{Oh2013}, beam wavefront  \cite{Kuk2016,Zhang2018c,Sheng2021}, gas pressure and species  \cite{Yoo2019}, the spatiotemporal dynamics in plasma formation and pulse propagation have been modified to apparently influence THz radiation. 

In the two-color scheme, the temporal and spatial overlapping of $\omega$ and $2\omega$ beams is a basic premise in beam alignment, where the two plasmas induced by the $\omega$ and $2\omega$ beams should spatially merge into a big plasma (overlapped plasma) acting as the best experimental condition. However, we found, when the two-color plasmas are concentrically separated along the propagation axis into cascading plasmas, the characteristics of THz radiation is significantly improved. Therefore, we introduce a new control knob, the distance between the two-color foci, to considerably increase the yield and bandwidth of THz emission. In this paper, the THz strength and spectral profile generated by linearly and circularly-polarized two-color fields, are measured as a function of the distance, which shows that the maximum conversion efficiency exceeds $10^{-3}$ with bandwidth $>$100 THz by using 800 nm, 35 fs driving lasers. The theory involving laser pulse propagation equation and photocurrent model indicates that the spatial redistribution of laser intensity in cascading plasmas balances THz generation and absorption in the plasma channel, leading to the maximum THz output.

\section{Experiment}\label{sec2}

\begin{figure}[h]%
\centering
\includegraphics[width=0.9\textwidth]{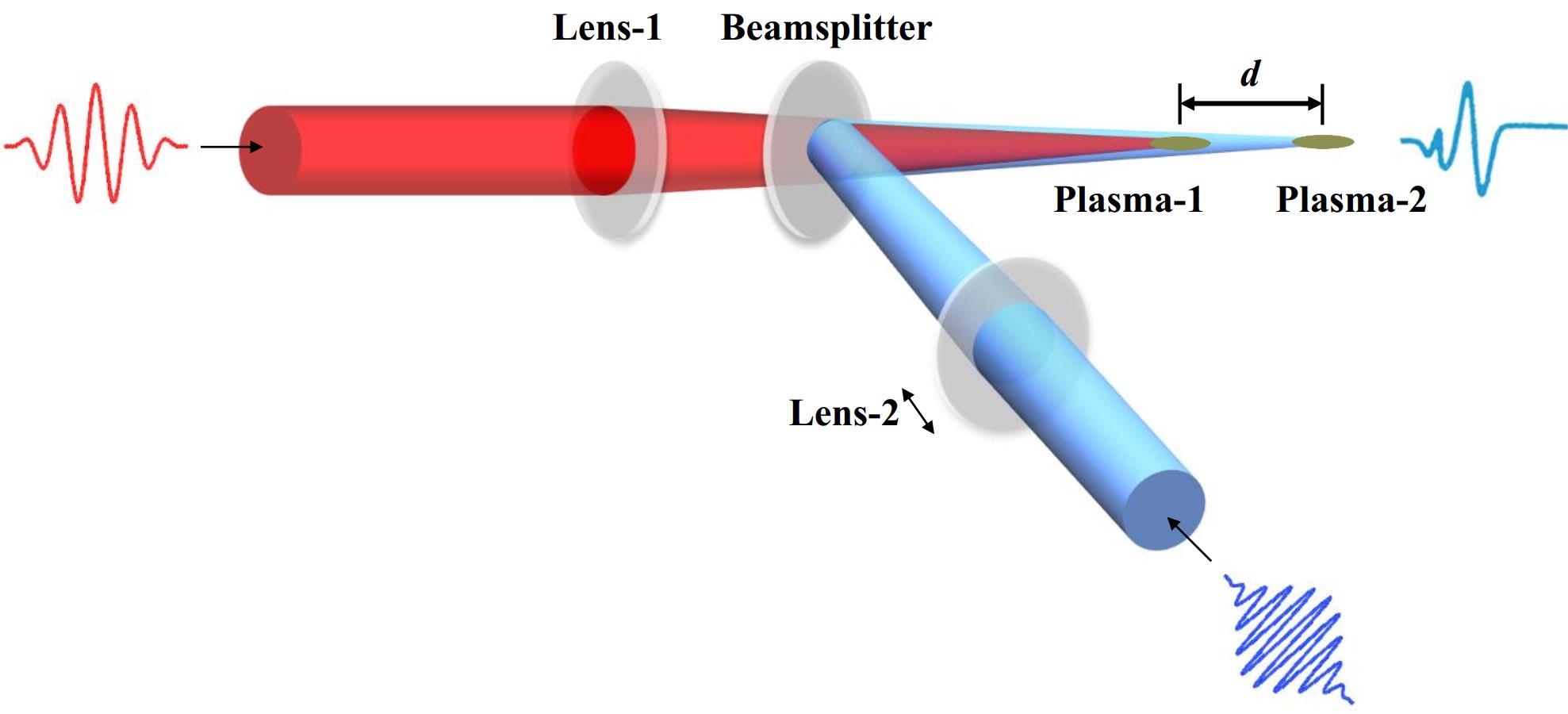}
\caption{Experimental schematics of ultra-broadband THz amplification. The $\omega$ and 2$\omega$ beams are respectively focused by two identical lenses with 10 cm focus length. In the experiment, we fix the lens position of $\omega$ beam, and the position of the 2$\omega$ lens is longitudinally moved along the propagation direction. The profile of the plasma channel strongly depends on the displacement $d$ of the 2$\omega$ lens, which is introduced as a new knob to amplify THz radiation.}\label{fig1}
\end{figure}

The experiment is implemented on a Ti:sapphire femtosecond amplification system. The $\omega$ beam with center wavelength of 800 nm is converted to the 2$\omega$ beam by a $\beta$-barium borate (BBO) crystal. The $\omega$ and 2$\omega$ beams respectively pass through two arms of a Michelson interferometer, and the polarization of the two beams can be individually controlled by the waveplates in the two arms. The lengths of the two arms are actively stabilized, and the relative phase delay $\tau$ of the $\omega$-2$\omega$ pulses is tunable with the accuracy of sub-femtosecond. The commonly-used focusing geometry, using one mirror (or lens) to focus two-color beams, is not applied in our setup, whereas a complicated focusing geometry is utilized. We place two identical lenses with 10 cm focus length in two arms of the interferometer to respectively focus the $\omega$ and 2$\omega$ beams, and the position of the lens can be moved along the propagation direction to change the focusing condition and plasma channel profile. As shown in Fig.~\ref{fig1}, the foci of the $\omega$ and 2$\omega$ beams are spatially separated to form cascading plasmas. The strength and bandwidth of the THz radiation strongly depends on the distance between the foci of $\omega$ and 2$\omega$ beams, which is controlled by the lens displacement $d$ of the 2$\omega$ beam. The THz strength as a function of $d$ is calibrated with electro-optical sampling (EOS) and pyroelectric detector, and the bandwidth is measured with a Fourier transform spectroscopy. The setup of THz generation and detection is referred to \textit{Method Section}.

\section{Experimental Results}\label{sec3}

\begin{figure}
\centering
\includegraphics[width=0.8\textwidth]{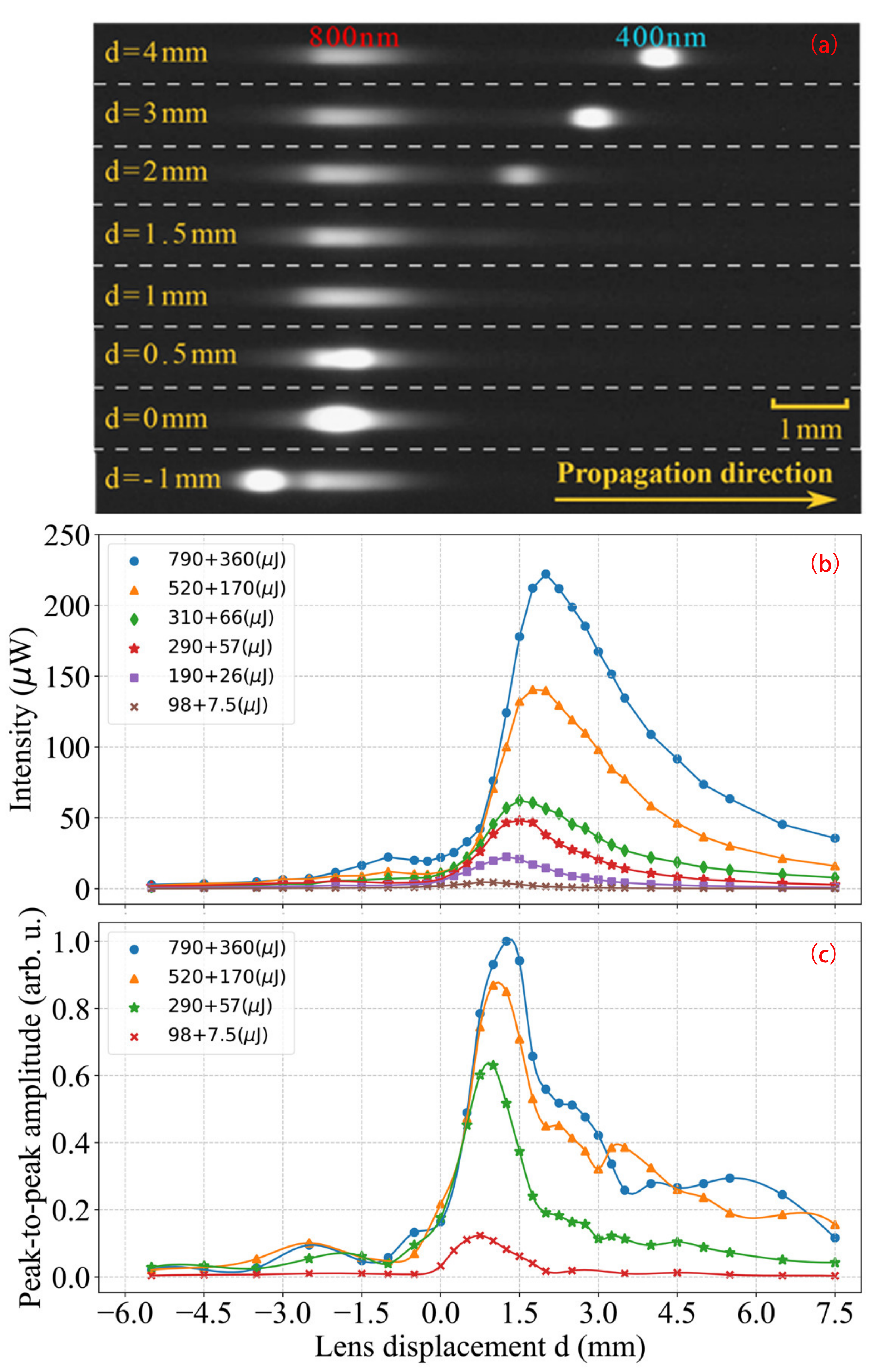}
\caption{Terahertz amplification in cascading plasmas. (a) The profile of cascading plasmas versus lens displacement $d$. The $\omega$ and 2$\omega$ beams produce the plasmas respectively. The two plasmas are spatially overlapped when $d=0 \ \mathrm{mm}$, and concentrically separated when $d$ is varied.  The plasmas produced by 400 nm and 800 nm beams are marked. (b) The THz power versus $d$ and input laser pulse energy $I_{\mathrm{THz}} (d, I_{\omega},I_{\mathrm{2}\omega})$ measured with pyroelectric detector. The \textit{x} axis is lens displacement $d$, and \textit{y} axis is THz power. The THz pulse is generated by the co-rotating circularly-polarized two-color fields. Surprisingly, the cascading plasmas at $d=2\ \mathrm{mm}$ radiates the most intense THz pulse, which is significantly larger than that at $d=0\ \mathrm{mm}$. (c) The \textit{s}-polarized THz electric field versus $d$ and input laser pulse energy $\boldsymbol{E}_{\mathrm{THz}} (d, I_{\omega},I_{\mathrm{2}\omega})$, which is measured with electro-optical sampling.} 
\label{fig2}
\end{figure}

Firstly, we measured the THz intensity $I_{\mathrm{THz}}$ versus the lens displacement $d$. In our setup, both the $\omega$ and 2$\omega$ beams produce plasmas, and the distance between the plasmas can be changed. The plasma profiles versus $d$ is recorded by a CCD camera, as shown in Fig.~\ref{fig2}(a). In commonly used setups, two plasmas are spatially overlapped to form a single plasma, where we define as $d=0\ \mathrm{mm}$. The third harmonic generation \cite{He2021} versus $d$ is used to precisely calibrate the position $d=0\ \mathrm{mm}$.  When changing $d$ the big plasma is separated along the propagation direction into two cascading plasmas, and the distance between the cascading plasmas approximately equals the lens displacement $d$. The brightness of 2$\omega$ plasma is obviously varied versus $d$. When the 2$\omega$ plasma approaches the $\omega$ plasma, the brightness of 2$\omega$ plasma suddenly decreases, reflecting the complex process in plasma channel formation.

After the spatiotemporal optimization of the two-color beams, the THz power versus $d$ and input laser power $I_{\mathrm{THz}}(d, I_{\omega},I_{\mathrm{2}\omega})$ is calibrated by a commercial pyroelectric detector, shown in Fig.~\ref{fig2}(b). Here, the co-rotating circularly-polarized two-color fields are used to obtain maximum conversion efficiency, and the THz emissions $I_{\mathrm{THz}}(d)$ in linearly polarized two-color fields have the similar $d$-dependent behavior. It is surprising that the THz radiation is significantly amplified by over one order of magnitude at $d=2\ \mathrm{mm}$ when the two plasmas are spatially well separated along propagation direction. The maximum conversion efficiency is obtained when the plasma formed by $\omega$ field is spatially ahead of the 2$\omega$ plasma. When decreasing input laser power, the maximum $I_{\mathrm{THz}}(d, I_{\omega},I_{\mathrm{2}\omega})$ approaches $d=0\ \mathrm{mm}$. The observation conflicts with the empirical awareness that the spatial overlapping of two-color foci is a prerequisite for the most efficient terahertz generation. 

Here, we can approximately estimate the conversion efficiency in cascade plasmas. Before the pyroelectric detector, two silicon wafers block the scattering light with $\sim$0.2 transmittance of THz beam. And the chopper blocks half of the input laser energy. Thus, the THz pulse energy at $d=2\ \mathrm{mm}$ exceeds 1 $\mu$J, and the conversion efficiency approaches 0.002, which is one order of magnitude higher than the analogous configuration with efficiency of $10^{-4}$. The estimation can also be confirmed in Fig.~\ref{fig2}(b), where $I_{\mathrm{THz}}(d=2\ \mathrm{mm})$ is $\sim$10 times higher than $I_{\mathrm{THz}}(d=0\ \mathrm{mm})$.

To further confirm the amplification in cascading plasmas, the THz electric field $\ensuremath{\boldsymbol{E}}_{\mathrm{THz}}(d, I_{\omega},I_{\mathrm{2}\omega})$ is measured with EOS, shown in Fig.~\ref{fig2}(c), which exhibits the similar behavior as $I_{\mathrm{THz}}(d, I_{\omega},I_{\mathrm{2}\omega})$. Here, only \textit{s}-polarized $\ensuremath{\boldsymbol{E}}_{\mathrm{THz}}$ is shown in the main text, and \textit{p}-polarized $\ensuremath{\boldsymbol{E}}_{\mathrm{THz}}$ has a similar manner (\textit{Supplementary Information}). Comparing to the intensity measurement with pyroelectric detector, the EOS measurement shows two different manners: (1) The $\ensuremath{\boldsymbol{E}}_{\mathrm{THz}}$ is sensitive to phase delay $\tau$ of $\omega$-2$\omega$ fields, but $I_{\mathrm{THz}}$ is not $\tau$ dependent. By scanning $\tau$, $\ensuremath{\boldsymbol{E}}_{\mathrm{THz}}$ is periodically modulated versus $\tau$, which is presented in \textit{Supplementary Information}. (2) The maximum efficiencies appear at different $d$. The maximum $I_{\mathrm{THz}}$ appears at $d=2\ \mathrm{mm}$, whereas $\ensuremath{\boldsymbol{E}}_{\mathrm{THz}}$ is optimized at $d=1.25\ \mathrm{mm}$. The discrepancy probably originates from the detection bandwidth of the two methods. The EOS only has response below 3 THz, while the pyroelectric detector has linear and relatively flat response function in the range of terahertz to infrared frequency. A possible explanation for unpronounced $\tau$-dependent $I_{\mathrm{THz}}$ is that, the high-frequency contributions from different spatial positions are averaged in the far field, which washes out the yield modulation versus $\tau$.

\begin{figure}
\centering
\includegraphics[width=0.8\textwidth]{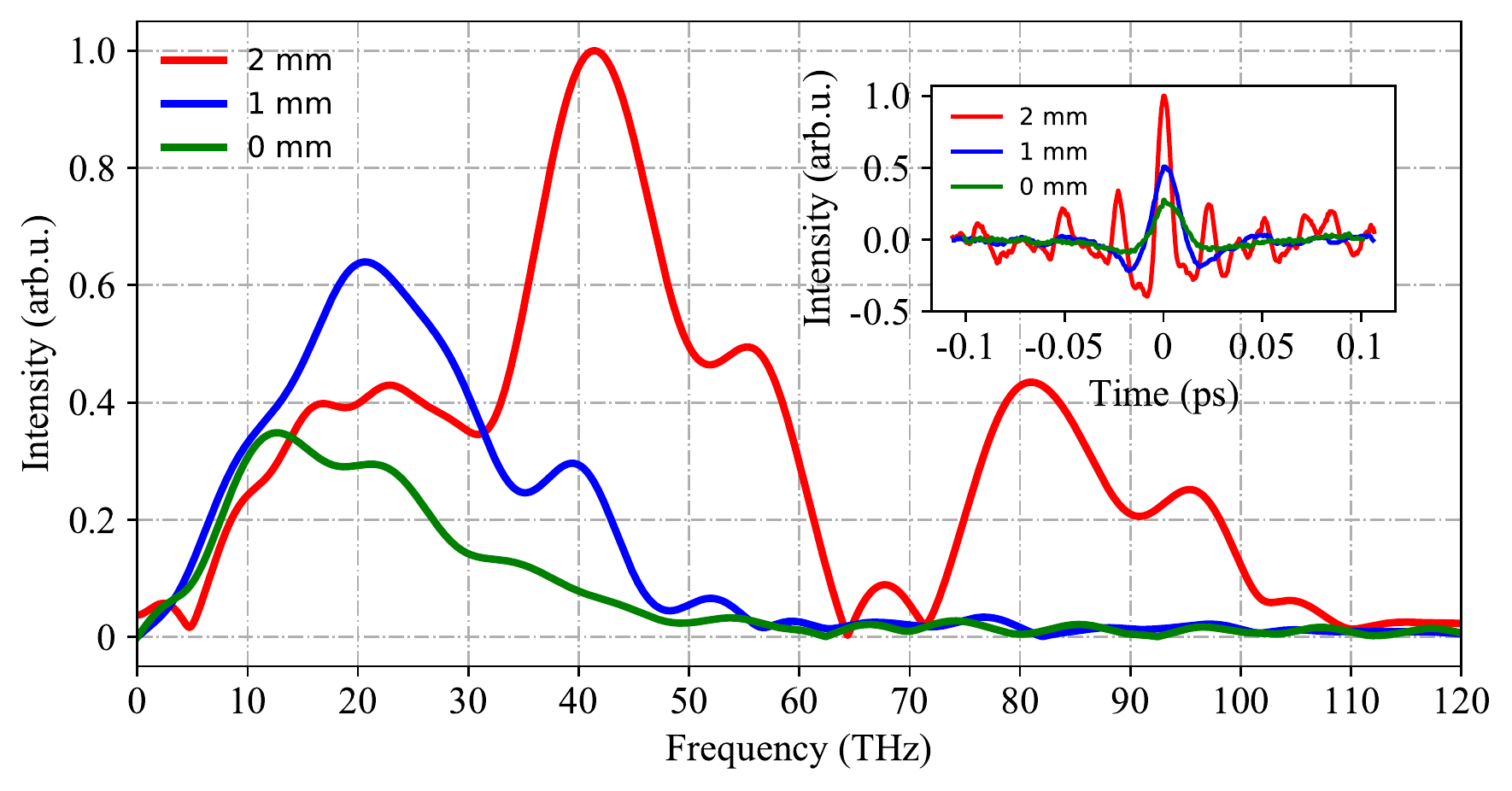}
\caption{Terahertz bandwidth versus $d$ between two-color plasmas. The spectral features of THz generation $\tilde{E}(\nu)$ in linearly polarized two-color fields are presented at $d=-2\ \mathrm{mm}$ (red), $d=-1\ \mathrm{mm}$ (blue) and $d=0\ \mathrm{mm}$ (green). The spectral intensities are represented on linear scale. When separating the two-color plasmas, $\tilde{E}(\nu)$ is significantly broadened and shifted to high frequency region. Inset: The interferograms obtained by Fourier transform spectroscopy.}
\label{fig3}
\end{figure}

The THz spectral features $\tilde{I}_{\mathrm{THz}}(\nu)$ at $I_{\omega} = 870\ \mathrm{\mu J},I_{\mathrm{2}\omega} = 460\ \mathrm{\mu J}$ are measured with Fourier transform spectroscopy, shown in Fig.~\ref{fig3}. The home-built Fourier transform spectrometer has been calibrated with a commercial optical parametric amplifier. Due to the noise, the very low-frequency region ($< 5$ THz) is not credible, which however can be measured with EOS. The spectral measurement shows that, when varying $d$ between two-color plasmas, the THz bandwidth is significantly broadened above 100 THz. The bandwidth broadening can be validated by the interferograms (Fig.~\ref{fig3} Inset). The temporal cycles emitted from cascading plasmas is much narrower than that in overlapped plasma. 
The spectrum measurement exhibits that, our new method can not only boost the THz strength, which can achieve comparable conversion efficiency as strong terahertz generation by tilted wave front excitation in lithium niobate crystal, but can also span THz spectral bandwidth up to mid-infrared region. 

\section{Theory}\label{sec4}

In order to understand the conversion efficiency enhancement in cascading plasmas, the THz radiation is numerically investigated with a (2D+1) laser pulse propagation equation combined with photocurrent model. In simulation, the free electron ensemble in cascading plasmas accelerated by asymmetric $\omega$-2$\omega$ fields induces residual photocurrent, which leads to the THz radiation. The absorption of the THz wave in plasma channel is also included in the model (\textit{Method Section}).

\begin{figure}
\centering
\includegraphics[width=0.75\textwidth]{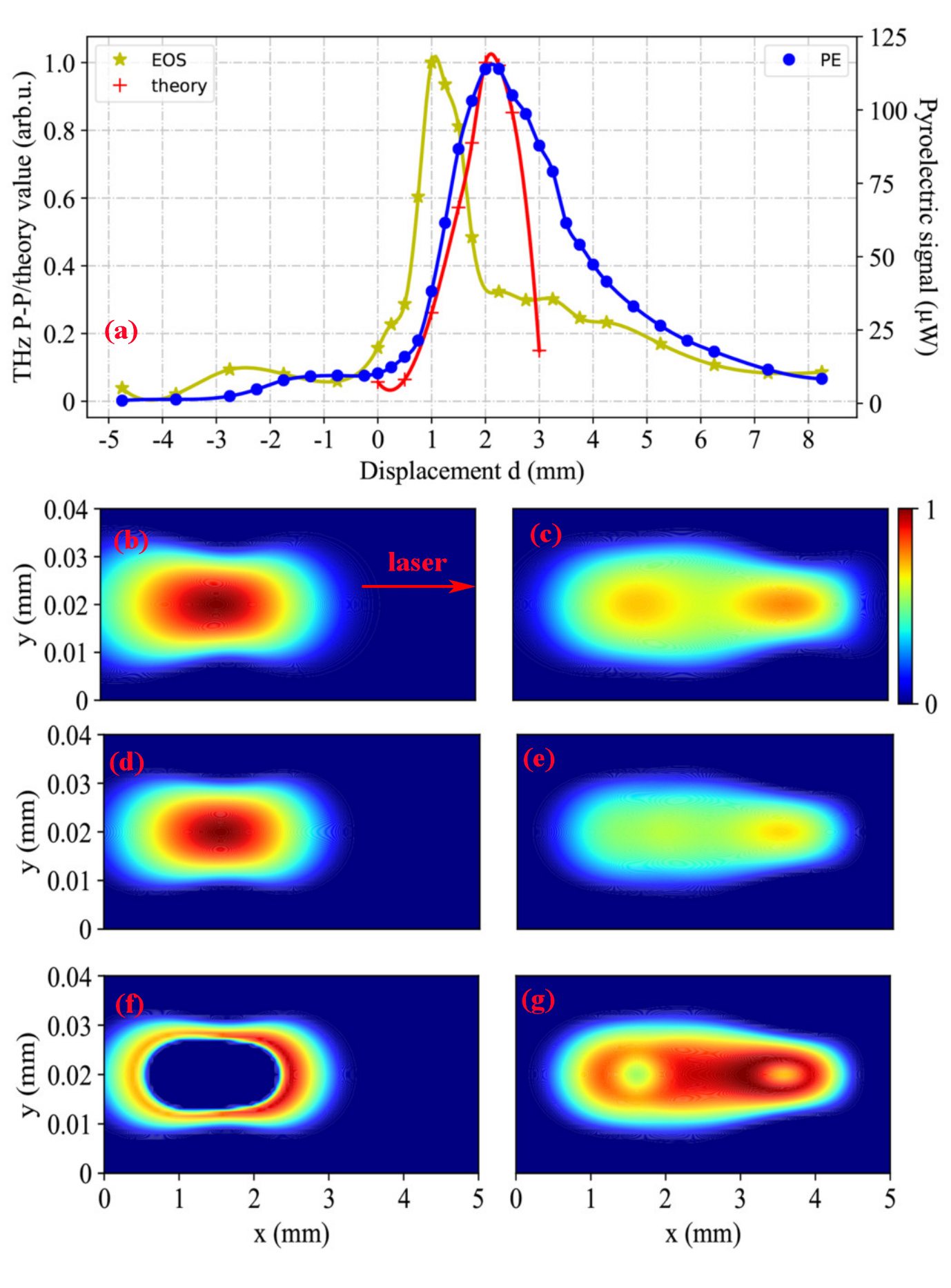}
\caption{Comparison between measurement and theory. (a) THz generation $I_{\mathrm{THz}}(d)$ (blue dots) and  $\ensuremath{\boldsymbol{E}}_{\mathrm{THz}}(d)$ (yellow stars) when $I_{\omega} = 870\ \mathrm{\mu J},I_{\mathrm{2}\omega} = 460\ \mathrm{\mu J}$ in linearly polarized $\omega$-2$\omega$ fields, and theoretical prediction (red line). (b) (c) Simulated spatial distributions of electron density in plasma at $d=0\ \mathrm{mm}$ and $d= 2\ \mathrm{mm}$. In panel (c) at $d= 2\ \mathrm{mm}$, the $\omega$ and 2$\omega$ foci are spatially separated into cascading plasmas. The left plasma is formed by the $\omega$ beam, and the right plasma is formed by the 2$\omega$ beam. The simulation well reproduces the fluorescence images of plasmas in Fig.~\ref{fig2} (a), which reflect electron density distributions in plasmas. (d) (e) The theoretical results of spatial distributions of THz generation. (f) (g) THz generation minus absorption (net THz emission). The dense plasma obviously attenuates THz emission. All the values in contour maps are logarithmically scaled and normalized to the maximum.}
\label{fig4}
\end{figure}

The THz radiation $I_{\mathrm{THz}}(d)$ and  $\ensuremath{\boldsymbol{E}}_{\mathrm{THz}}(d)$ at $I_{\omega} = 870\ \mathrm{\mu J},I_{\mathrm{2}\omega} = 460\ \mathrm{\mu J}$ in linearly polarized $\omega$-2$\omega$ fields are compared to the theoretical simulation, as shown in Fig.~\ref{fig4}(a). In simulation, we estimate that the waists of $\omega$ and 2$\omega$ beams at the foci are $20\ \mathrm{\mu m}$ and $10\ \mathrm{\mu m}$, and peak powers of $\omega$ and 2$\omega$ fields are $100\ \mathrm{TW/cm^2}$ and $150\ \mathrm{TW/cm^2}$. 
Fig.~\ref{fig4}(a) shows the simulated THz strength versus $d$ at 1 THz, which approximately agrees with the measured $d$-dependent behaviors of $I_{\mathrm{THz}}(d)$ and  $\ensuremath{\boldsymbol{E}}_{\mathrm{THz}}(d)$.

We investigate electron density distribution, THz generation and absorption in space, and give a preliminary explanation for the THz enhancement in cascading plasmas. As shown in Fig.~\ref{fig4}(b) and (c), when the foci of $\omega$ and 2$\omega$ spatially overlap at $d=0\ \mathrm{mm}$, the two-color fields produce very dense plasma, which is confined to a small volume with high electron density and strong gradient.
In the center of the plasma volume, the electron density is estimated as $n_e \sim 10^{18}\ /\mathrm{cm}^3$.
Comparatively, when the two-color foci are separated, the plasma is stretched into cascading plasmas with larger length and lower electron density. The electron density in cascade plasmas $n_e \sim 10^{16}\ /\mathrm{cm}^3$, which is  more homogeneously spatially distributed than the overlapped plasma.
 
The THz generation is highly relevant to electron density. Although the spatial density of THz emission in the cascading plasmas (Fig.~\ref{fig4}(e)) is smaller than that in overlapped plasma (Fig.~\ref{fig4}(d)) due to low electron density, yet, it is compensated after full space integration of plasma volume. Therefore, the overlapped plasma and cascading plasmas have similar throughput of THz emission. It can be further verified by the residual photocurrents at spatial samples at $z = 1.5,\  2.5,\  3.5\ \mathrm{mm}$  (\textit{Supplementary Information}), indicating the spatial THz generation in plasma. 

The net THz emission from plasma depends on both THz generation and absorption in plasma. Here, we define absorption length $L_a$ to describe how long the THz wave is able to propagate in plasma filament. In the center of the overlapped plasma, the electron density $n_e \sim 10^{18} /\mathrm{cm}^3$, corresponds to absorption length $L_a \sim 1\ \mathrm{\mu m}$, which is far less than the plasma length. Hence, the THz wave is mostly depleted in plasma volume, as shown in (Fig.~\ref{fig4}(f)). Comparatively, the absorption length in cascading plasmas can be estimated as $L_a \sim 6.8\ \mathrm{mm}$ according to average electron density of plasma volume. It indicates that the THz absorption in cascading plasmas is much weaker than that in overlapped plasma (Fig.~\ref{fig4}(e)). Thus, by manipulating spatial redistribution of laser input energy, the THz generation and absorption are self optimized in cascading plasmas, finally achieving one order of magnitude enhancement of conversion efficiency.

The THz spectral broadening can hardly be explained with the plasma absorption. We tentatively attribute spectral broadening to spatiotemporal reshaping of driving fields when the ultrashort pulse propagates in cascading plasmas. The dispersion and highly nonlinear effects in plasma filament lead to the temporal distortion and compression of two-color fields. The pulse reshaping would induce the rapidly time-varying photocurrent, leading to high-frequency THz component. Since the nonlinear interaction length in cascading plasmas is longer than the length of the overlapped plasma, the spectral broadening in cascading plasmas is more pronounced than that in overlapped plasma. More convincing explanations need further theoretical investigation, which may involve highly complex and rich spatiotemporal dynamics during pulse propagation in plasma filament.

\section{Conclusion}\label{sec5}

We introduce a new control knob, the distance between the two-color cascading plasmas, to promote THz radiation in two-color scheme. With the distance appropriately optimized, the THz conversion efficiency reaches $\sim10^{-3}$ and the bandwidth can be broadened $>$100 THz with 800 nm, 35 fs laser pulse. The conversion efficiency is one order of magnitude higher and the bandwidth is more than 2 times broader than the counterpart configuration of overlapped plasma. The new proposed geometry can achieve considerable brightness and supercontinuum bandwidth with fairly simple setup, which avoids sophisticated optical parametric amplifier and pulse compression technique. 

The first-ever proposed method may also be applicable in long wavelength driving THz generation for further enhancement conversion efficiency, which may break the current records of the strength and bandwidth of THz ultrashort pulses. The ultra-broadband feature of THz radiation has potential applicability to study structure and ultrafast dynamics of complex systems, whereas the strength feature can be applied in nonlinear optics, strong field physics and accelerator technique. The propagation equation combined with photocurrent model indicates that the THz amplification originates from the interplay between THz generation and plasma absorption in cascading plasmas. The complex mechanism of THz radiation in cascade plasma channel is still an open question calling for further theoretical exploration.

\section{Methods}
\subsection{Experimental setup}
The experimental setup is shown in \textit{Supplementary Information}. A Ti:sapphire laser delivers \textit{p}-polarized femtosecond pulse with 35 fs, $\sim$1.8 mJ/pulse, centered at 800 nm with 60 nm bandwidth. 
The $\omega$ beam passes through a 200 $\mu$m type-I {$\beta$}-barium borate (BBO) crystal, and a \textit{s}-polarized 2$\omega$ beam is generated (conversion efficiency $\sim$30\%). The co-propagating $\omega$-2$\omega$ beams are separated by a dichroic mirror (DM-2) into the two arms of a Michelson interferometry. The polarization of the $\omega$-2$\omega$ beams can be arbitrarily controlled by quarter-waveplates. 

The phase jitter between $\omega$-2$\omega$ beams are suppressed by introducing a actively stabilized Michelson interferometry. To stabilize the relative phase of two arms of Michelson interferometry, a continuous green laser (532 nm) co-propagates with the $\omega$-2$\omega$ beams and interferes. The interference fringes are monitored by a CCD camera as a feedback signal. A mirror fixed on a piezo actuator provides a real-time feedback to keep the interference fringes stable. 
After stabilization, the relative phase fluctuation in the system is smaller than $0.02\pi$ during data acquisition. Due to the difference of refractive indices between the $\omega$-2$\omega$ fields in air, the phase delay $\tau$ can be tuned with sub-femtosecond accuracy by changing the distance between the BBO and air plasma.

The $\omega$-2$\omega$ beams are respectively focused in two arms of the interferometry with L-1 and L-2 lenses with focal length of 10 cm. The L-2 lens is fixed, and L-1 lens is installed on a translation stage. The L-1 position $d$ can be moved along propagation direction. The $\omega$-2$\omega$ beams are combined with DM-3 and focused into atmospheric air to produce plasma. The plasma profile and THz conversion efficiency highly depend on $d$. And the plasma profile can be recorded by a CCD camera.

The THz strength is measured by two methods, electro-optical sampling (EOS) and pyroelectric detector. In EOS, a 800 $\mu$m thickness silicon wafer blocks the $\omega$ and 2$\omega$ beam. 
The THz beam is aligned by two parabolic mirrors and focused on a 1 mm thickness ZnTe crystal. A metal wire-grid polarizer filters out polarized THz beam, and the ZnTe crystal is fixed at the special orientation, which has the same responses for \textit{s}- and \textit{p}-polarized components. The EOS has detection bandwidth of $< 3$ THz and the polarization can be resolved. The THz intensity is also measured with a pyroelectric detector (THZ9B-BL-DZ, GENTEC-EO) with 25 Hz chopper frequency. Here, an 400 $\mu$m thickness silicon wafer is glued on the detector to block scattering light. The THz bandwidth is obtained with a home built Fourier transform spectrometer, where the pyroelectric detector is used as the detector. The wavelength calibration of the spectrometer in mid-infrared region is implemented with optical parametric amplifier.

\subsection{Theoretical Model}

The linearly polarized laser pulse propagation model consisting of the absorption loss of ionization, nonlinear Kerr effect and plasma defocusing is described as a three-dimensional (2D+1) Maxwell's wave equation \cite{Geissler1999}
\begin{equation}
\bigtriangledown ^2 E(r,z,t)-\frac{1}{c^2}\frac{\partial ^2 E(r,z,t)}{\partial t^2}=\mu_0 \frac{\partial J(r,z,t)}{\partial t}+\frac{\omega^2_0}{c^2}(1-\eta ^2_{\mathrm{eff}}) E(r,z,t), 
\end{equation}
where $E(r,z,t)$ denotes the transverse electric field at radial $r$ and axial $z$, and $\mu_0$, $\omega_0$ and $c$ are the permeability of vacuum, central frequency of electric field and speed of light in vacuum. In source terms, the absorption loss of ionization is given as \cite{Gaarde08}  $J(r,z,t)=\frac{W(t) n_{e}(t) I_p E(r,z,t)}{\vert E(r,z,t) \vert ^{2}}$, where $I_p$ , $W(t)$ and $n_e(t)$ are the ionization potential, ionization rate and free electron density. The effective refractive index in source term is written as $\eta_{\mathrm{eff}}=\eta_0 +\eta_2 I(r,z,t) - \omega^2_p /2\omega^2_0 $ , where the refraction and absorption of the neutral gas $\eta_0 \sim 1$, the nonlinear Kerr index in atmospheric air $\eta_2=3.2 \times 10^{-19} \mathrm{cm^2/W}$, the plasma defocusing $\omega^2_p / 2\omega^2_0 =n_e /2 n_c  $, $n_c[\mathrm{/cm^3}]\sim 1.1 \times 10^{21}/\lambda^2 [\mathrm{\mu m}]$ is the critical density of the laser with wavelength of $ \lambda $. The electron density of the photoionization induced plasma is calculated with the empirical Ammosov-Delone-Krainov (ADK) formula \cite{Tong05}
\begin{equation}
\frac{d n_e(t)}{dt}=W(t)n_a(t),
\label{eq:IonizationRate}
\end{equation} 
where $W(t)$ is the instantaneous tunnel ionization rate. When the first ionization dominates, 
the time-dependent neutral density is written as $n_a(t)=n_0-n_e(t)$, $n_0$ is initial neutral gas density. As the major ingredient of air, the nitrogen is used to estimate the ionization rate of air. 

The transient photocurrent model \cite{Kre06} is employed to calculate THz generation in the two-color fields $E=E_\omega (r,z,t)+E_{2\omega}(r,z,t)$. The THz generation at a certain position is estimated as $E_{\mathrm{THz}} \propto \frac{dJ(t)}{dt}$. The transient current which is formed by  acceleration of free electrons in two-color fields can be expressed as 
\begin{equation}
J(t)=-e\int _ 0 ^ t \upsilon (t^{ \prime}, t) dn_e(t^{ \prime}).
\label{eq:NetCurrent}
\end{equation}
According to Eq. \ref{eq:IonizationRate}, the increment of electron density $dn_e$ depends on ionization rate. The transverse velocity that the electron acquires from $t^{\prime}$ to $t$ is given by
\begin{equation}
\upsilon (t^{\prime} ,t) = -\frac{e}{m}\int_{t^{\prime}}^t E(\xi)d\xi,
\label{eq:NetVelocity}
\end{equation}
where free electrons are assumed to be born with initial velocity $\upsilon(t^{\prime}) = 0$. The transverse velocity depends on the laser waveform between $t^{\prime}$ and $t$. 
According to Eq. \ref{eq:NetCurrent}, the residual current at the end of laser pulse can be expressed as $J(t)=-e\int _0 ^ \infty \upsilon_d (t^{ \prime}) dn_e(t^{ \prime})$. Considering Eq. \ref{eq:NetVelocity}, the transverse velocity at the end of laser pulse, i.e., the drift velocity, is written as $\upsilon_d (t^{\prime}) = -\frac{e}{m}\int_{t^{\prime}}^\infty E(\xi)d\xi$. According to integration formula, the drift velocity can be written as $\upsilon_d (t^{\prime}) = -\frac{e}{m}(\int_0 ^\infty E(\xi)d\xi-\int_0 ^{t^{\prime}} E(\xi)d\xi)=-\frac{e}{m}(A(\infty)-A(t^{\prime}))$. The two vector potentials $A(\infty)$ and $A(t^{\prime})$ are determined by asymmetry of the two-color fields. 

When THz wave propagates in a dense plasma, the THz emission with frequency $\Omega$ depends on absorption length of the plasma medium. Because the coherence length between $\omega$ and $2\omega$ is far longer than absorption length for gaseous plasma, the absorption effect dominates \cite{Constant1999}.
The THz intensity without the dependence of phase shift is calculated by
\begin{equation}
I_{\mathrm{THz}}=\vert \iint E_{\mathrm{THz}}(\Omega)\exp( -\frac{L-z}{L_a} ) drdz \vert ^2,
\end{equation}
where the $L$ refers to plasma length. The absorption length depends on the electron–ion collisional frequency $\nu$ and the plasma frequency $\omega_p$, which is written as $L_a\sim 2c(\Omega^2+\nu^2)/(\omega^2_p \nu)$. The electron–ion collisional frequency is estimated as $\nu \sim 2.9 \times 10^{-6}Z^2N_i [\mathrm{/cm^3}] \ln{\Lambda _{ei}} (T_{\mathrm{eff}}[\mathrm{eV}])^{-3/2}$, 
where $N_i$ is the ion density, $\ln {\Lambda}$ is the Coulomb logarithm, the effective temperature $T_{\mathrm{eff}}=\kappa_{B}T_{e}+2U_p/3$. The ponderomotive potential is written as $U_p=\frac{1}{4}\frac{e^2 E^2_0}{m \omega^2_0} \sim 9.33 \times 10^{-14} I_0 (\mathrm{W/cm^2}) \lambda^2_0 (\mu m) $. In the simulation, the local fluctuation of electron temperature is neglected, and the average temperature $T_e$ is given by an average electron–ion collisional frequency $\tilde{\nu} \sim 3 \times 10^{12}\ \mathrm{Hz}$ in the gaseous plasma when laser intensity is $\sim 100\ \mathrm{TW/cm^2}$. In the inverse bremsstrahlung heating regime, when the heating time scale $\tau^{\ast}=(1/\tilde{\nu})^{3/10} \sim 90\  \mathrm{fs} $  is larger than the 
pulse duration $\tau \sim 35\ \mathrm{fs}$, the electron temperature in short-pulse regime \cite{Durfee1995} is estimated by $\kappa_{B}T_{e} \sim 2U_p \tau/(5\tau^{\ast}) \sim 1.7\ \mathrm{eV}$.

\bmhead{Acknowledgments}
The work is supported by National Natural Science Foundation of China (NSFC) (12174284, 11827806, 11874368, 11864037, 91850209). We also acknowledge the support from Shanghai-XFEL beamline project (SBP) and Shanghai High repetition rate XFEL and Extreme light facility (SHINE).

\bibliography{Main}


\begin{thebibliography}{37}
\ifx \bisbn   \undefined \def \bisbn  #1{ISBN #1}\fi
\ifx \binits  \undefined \def \binits#1{#1}\fi
\ifx \bauthor  \undefined \def \bauthor#1{#1}\fi
\ifx \batitle  \undefined \def \batitle#1{#1}\fi
\ifx \bjtitle  \undefined \def \bjtitle#1{#1}\fi
\ifx \bvolume  \undefined \def \bvolume#1{\textbf{#1}}\fi
\ifx \byear  \undefined \def \byear#1{#1}\fi
\ifx \bissue  \undefined \def \bissue#1{#1}\fi
\ifx \bfpage  \undefined \def \bfpage#1{#1}\fi
\ifx \blpage  \undefined \def \blpage #1{#1}\fi
\ifx \burl  \undefined \def \burl#1{\textsf{#1}}\fi
\ifx \doiurl  \undefined \def \doiurl#1{\url{https://doi.org/#1}}\fi
\ifx \betal  \undefined \def \betal{\textit{et al.}}\fi
\ifx \binstitute  \undefined \def \binstitute#1{#1}\fi
\ifx \binstitutionaled  \undefined \def \binstitutionaled#1{#1}\fi
\ifx \bctitle  \undefined \def \bctitle#1{#1}\fi
\ifx \beditor  \undefined \def \beditor#1{#1}\fi
\ifx \bpublisher  \undefined \def \bpublisher#1{#1}\fi
\ifx \bbtitle  \undefined \def \bbtitle#1{#1}\fi
\ifx \bedition  \undefined \def \bedition#1{#1}\fi
\ifx \bseriesno  \undefined \def \bseriesno#1{#1}\fi
\ifx \blocation  \undefined \def \blocation#1{#1}\fi
\ifx \bsertitle  \undefined \def \bsertitle#1{#1}\fi
\ifx \bsnm \undefined \def \bsnm#1{#1}\fi
\ifx \bsuffix \undefined \def \bsuffix#1{#1}\fi
\ifx \bparticle \undefined \def \bparticle#1{#1}\fi
\ifx \barticle \undefined \def \barticle#1{#1}\fi
\bibcommenthead
\ifx \bconfdate \undefined \def \bconfdate #1{#1}\fi
\ifx \botherref \undefined \def \botherref #1{#1}\fi
\ifx \url \undefined \def \url#1{\textsf{#1}}\fi
\ifx \bchapter \undefined \def \bchapter#1{#1}\fi
\ifx \bbook \undefined \def \bbook#1{#1}\fi
\ifx \bcomment \undefined \def \bcomment#1{#1}\fi
\ifx \oauthor \undefined \def \oauthor#1{#1}\fi
\ifx \citeauthoryear \undefined \def \citeauthoryear#1{#1}\fi
\ifx \endbibitem  \undefined \def \endbibitem {}\fi
\ifx \bconflocation  \undefined \def \bconflocation#1{#1}\fi
\ifx \arxivurl  \undefined \def \arxivurl#1{\textsf{#1}}\fi
\csname PreBibitemsHook\endcsname

\bibitem{Cook00}
\begin{barticle}
\bauthor{\bsnm{Cook}, \binits{D.J.}},
\bauthor{\bsnm{Hochstrasser}, \binits{R.M.}}:
\batitle{Intense terahertz pulses by four-wave rectification in air}.
\bjtitle{Opt. Lett.}
\bvolume{25}(\bissue{16}),
\bfpage{1210}--\blpage{1212}
(\byear{2000}).
\doiurl{10.1364/OL.25.001210}
\end{barticle}
\endbibitem

\bibitem{Koulouklidis2020}
\begin{barticle}
\bauthor{\bsnm{Koulouklidis}, \binits{A.D.}},
\bauthor{\bsnm{Gollner}, \binits{C.}},
\bauthor{\bsnm{Shumakova}, \binits{V.}},
\bauthor{\bsnm{Fedorov}, \binits{V.Y.}},
\bauthor{\bsnm{Pug{\v{z}}lys}, \binits{A.}},
\bauthor{\bsnm{Baltu{\v{s}}ka}, \binits{A.}},
\bauthor{\bsnm{Tzortzakis}, \binits{S.}}:
\batitle{{Observation of extremely efficient terahertz generation from
  mid-infrared two-color laser filaments}}.
\bjtitle{Nature Communications}
\bvolume{11}(\bissue{1}),
\bfpage{292}
(\byear{2020}).
\doiurl{10.1038/s41467-019-14206-x}
\end{barticle}
\endbibitem

\bibitem{Mitrofanov2020}
\begin{barticle}
\bauthor{\bsnm{Mitrofanov}, \binits{A.V.}},
\bauthor{\bsnm{Sidorov-Biryukov}, \binits{D.A.}},
\bauthor{\bsnm{Nazarov}, \binits{M.M.}},
\bauthor{\bsnm{Voronin}, \binits{A.A.}},
\bauthor{\bsnm{Rozhko}, \binits{M.V.}},
\bauthor{\bsnm{Shutov}, \binits{A.D.}},
\bauthor{\bsnm{Ryabchuk}, \binits{S.V.}},
\bauthor{\bsnm{Serebryannikov}, \binits{E.E.}},
\bauthor{\bsnm{Fedotov}, \binits{A.B.}},
\bauthor{\bsnm{Zheltikov}, \binits{A.M.}}:
\batitle{{Ultraviolet-to-millimeter-band supercontinua driven by ultrashort
  mid-infrared laser pulses}}.
\bjtitle{OPTICA}
\bvolume{7}(\bissue{1}),
\bfpage{15}--\blpage{19}
(\byear{2020}).
\doiurl{10.1364/OPTICA.7.000015}
\end{barticle}
\endbibitem

\bibitem{Jang2019}
\begin{barticle}
\bauthor{\bsnm{Jang}, \binits{D.}},
\bauthor{\bsnm{Schwartz}, \binits{R.M.}},
\bauthor{\bsnm{Woodbury}, \binits{D.}},
\bauthor{\bsnm{Griff-McMahon}, \binits{J.}},
\bauthor{\bsnm{Younis}, \binits{A.H.}},
\bauthor{\bsnm{Milchberg}, \binits{H.M.}},
\bauthor{\bsnm{Kim}, \binits{K.-Y.}}:
\batitle{{Efficient terahertz and Brunel harmonic generation from air plasma
  via mid-infrared coherent control}}.
\bjtitle{OPTICA}
\bvolume{6}(\bissue{10}),
\bfpage{1338}--\blpage{1341}
(\byear{2019}).
\doiurl{10.1364/OPTICA.6.001338}
\end{barticle}
\endbibitem

\bibitem{Pashkin2011}
\begin{barticle}
\bauthor{\bsnm{Pashkin}, \binits{A.}},
\bauthor{\bsnm{K\"ubler}, \binits{C.}},
\bauthor{\bsnm{Ehrke}, \binits{H.}},
\bauthor{\bsnm{Lopez}, \binits{R.}},
\bauthor{\bsnm{Halabica}, \binits{A.}},
\bauthor{\bsnm{Haglund}, \binits{R.F.}},
\bauthor{\bsnm{Huber}, \binits{R.}},
\bauthor{\bsnm{Leitenstorfer}, \binits{A.}}:
\batitle{Ultrafast insulator-metal phase transition in vo${}_{2}$ studied by
  multiterahertz spectroscopy}.
\bjtitle{Phys. Rev. B}
\bvolume{83},
\bfpage{195120}
(\byear{2011}).
\doiurl{10.1103/PhysRevB.83.195120}
\end{barticle}
\endbibitem

\bibitem{Valverde-Chavez2015}
\begin{barticle}
\bauthor{\bsnm{Valverde-Ch{\'{a}}vez}, \binits{D.A.}},
\bauthor{\bsnm{Ponseca}, \binits{C.S.}},
\bauthor{\bsnm{Stoumpos}, \binits{C.C.}},
\bauthor{\bsnm{Yartsev}, \binits{A.}},
\bauthor{\bsnm{Kanatzidis}, \binits{M.G.}},
\bauthor{\bsnm{Sundstr{\"{o}}m}, \binits{V.}},
\bauthor{\bsnm{Cooke}, \binits{D.G.}}:
\batitle{{Intrinsic femtosecond charge generation dynamics in single crystal
  CH3NH3PbI3}}.
\bjtitle{Energy and Environmental Science}
\bvolume{8}(\bissue{12}),
\bfpage{3700}--\blpage{3707}
(\byear{2015}).
\doiurl{10.1039/c5ee02503f}
\end{barticle}
\endbibitem

\bibitem{Wang2016}
\begin{barticle}
\bauthor{\bsnm{Wang}, \binits{T.}},
\bauthor{\bsnm{Romanova}, \binits{E.A.}},
\bauthor{\bsnm{Abdel-Moneim}, \binits{N.}},
\bauthor{\bsnm{Furniss}, \binits{D.}},
\bauthor{\bsnm{Loth}, \binits{A.}},
\bauthor{\bsnm{Tang}, \binits{Z.}},
\bauthor{\bsnm{Seddon}, \binits{A.}},
\bauthor{\bsnm{Benson}, \binits{T.}},
\bauthor{\bsnm{Lavrinenko}, \binits{A.}},
\bauthor{\bsnm{Jepsen}, \binits{P.U.}}:
\batitle{{Time-resolved terahertz spectroscopy of charge carrier dynamics in
  the chalcogenide glass As{\_}30Se{\_}30Te{\_}40 [Invited]}}.
\bjtitle{Photonics Research}
\bvolume{4}(\bissue{3}),
\bfpage{22}
(\byear{2016}).
\doiurl{10.1364/prj.4.000a22}
\end{barticle}
\endbibitem

\bibitem{Chen2016}
\begin{barticle}
\bauthor{\bsnm{Chen}, \binits{H.}},
\bauthor{\bsnm{Wen}, \binits{X.}},
\bauthor{\bsnm{Zhang}, \binits{J.}},
\bauthor{\bsnm{Wu}, \binits{T.}},
\bauthor{\bsnm{Gong}, \binits{Y.}},
\bauthor{\bsnm{Zhang}, \binits{X.}},
\bauthor{\bsnm{Yuan}, \binits{J.}},
\bauthor{\bsnm{Yi}, \binits{C.}},
\bauthor{\bsnm{Lou}, \binits{J.}},
\bauthor{\bsnm{Ajayan}, \binits{P.M.}},
\bauthor{\bsnm{Zhuang}, \binits{W.}},
\bauthor{\bsnm{Zhang}, \binits{G.}},
\bauthor{\bsnm{Zheng}, \binits{J.}}:
\batitle{{Ultrafast formation of interlayer hot excitons in atomically thin MoS
  2 /WS 2 heterostructures}}.
\bjtitle{Nature Communications}
\bvolume{7},
\bfpage{1}--\blpage{8}
(\byear{2016}).
\doiurl{10.1038/ncomms12512}
\end{barticle}
\endbibitem

\bibitem{Cossel17}
\begin{barticle}
\bauthor{\bsnm{Cossel}, \binits{K.C.}},
\bauthor{\bsnm{Waxman}, \binits{E.M.}},
\bauthor{\bsnm{Finneran}, \binits{I.A.}},
\bauthor{\bsnm{Blake}, \binits{G.A.}},
\bauthor{\bsnm{Ye}, \binits{J.}},
\bauthor{\bsnm{Newbury}, \binits{N.R.}}:
\batitle{Gas-phase broadband spectroscopy using active sources: progress,
  status, and applications (invited)}.
\bjtitle{J. Opt. Soc. Am. B}
\bvolume{34}(\bissue{1}),
\bfpage{104}--\blpage{129}
(\byear{2017}).
\doiurl{10.1364/JOSAB.34.000104}
\end{barticle}
\endbibitem

\bibitem{Zhang2018a}
\begin{barticle}
\bauthor{\bsnm{Zhang}, \binits{Y.}},
\bauthor{\bsnm{Yan}, \binits{T.-M.}},
\bauthor{\bsnm{Jiang}, \binits{Y.H.}}:
\batitle{{Ultrafast Mapping of Coherent Dynamics and Density Matrix
  Reconstruction in Terahertz-Assisted Laser Field.}}
\bjtitle{Physical Review Letters}
\bvolume{121}(\bissue{11}),
\bfpage{113201}
(\byear{2018}).
\doiurl{10.1103/PhysRevLett.121.113201}
\end{barticle}
\endbibitem

\bibitem{Zhang2018b}
\begin{barticle}
\bauthor{\bsnm{Zhang}, \binits{D.}},
\bauthor{\bsnm{Fallahi}, \binits{A.}},
\bauthor{\bsnm{Hemmer}, \binits{M.}},
\bauthor{\bsnm{Wu}, \binits{X.}},
\bauthor{\bsnm{Fakhari}, \binits{M.}},
\bauthor{\bsnm{Hua}, \binits{Y.}},
\bauthor{\bsnm{Cankaya}, \binits{H.}},
\bauthor{\bsnm{Calendron}, \binits{A.-L.}},
\bauthor{\bsnm{Zapata}, \binits{L.E.}},
\bauthor{\bsnm{Matlis}, \binits{N.H.}},
\bauthor{\bsnm{Kaertner}, \binits{F.X.}}:
\batitle{{Segmented terahertz electron accelerator and manipulator (STEAM)}}.
\bjtitle{Nature Photonics}
\bvolume{12}(\bissue{6}),
\bfpage{336}
(\byear{2018}).
\doiurl{10.1038/s41566-018-0138-z}
\end{barticle}
\endbibitem

\bibitem{Clerici2013}
\begin{barticle}
\bauthor{\bsnm{Clerici}, \binits{M.}},
\bauthor{\bsnm{Peccianti}, \binits{M.}},
\bauthor{\bsnm{Schmidt}, \binits{B.E.}},
\bauthor{\bsnm{Caspani}, \binits{L.}},
\bauthor{\bsnm{Shalaby}, \binits{M.}},
\bauthor{\bsnm{Gigu\`ere}, \binits{M.}},
\bauthor{\bsnm{Lotti}, \binits{A.}},
\bauthor{\bsnm{Couairon}, \binits{A.}},
\bauthor{\bsnm{L\'egar\'e}, \binits{F.m.c.}},
\bauthor{\bsnm{Ozaki}, \binits{T.}},
\bauthor{\bsnm{Faccio}, \binits{D.}},
\bauthor{\bsnm{Morandotti}, \binits{R.}}:
\batitle{Wavelength scaling of terahertz generation by gas ionization}.
\bjtitle{Phys. Rev. Lett.}
\bvolume{110},
\bfpage{253901}
(\byear{2013}).
\doiurl{10.1103/PhysRevLett.110.253901}
\end{barticle}
\endbibitem

\bibitem{Nguyen2019}
\begin{barticle}
\bauthor{\bsnm{Nguyen}, \binits{A.}},
\bauthor{\bsnm{Kaltenecker}, \binits{K.J.}},
\bauthor{\bsnm{Delagnes}, \binits{J.-C.}},
\bauthor{\bsnm{Zhou}, \binits{B.}},
\bauthor{\bsnm{Cormier}, \binits{E.}},
\bauthor{\bsnm{Fedorov}, \binits{N.}},
\bauthor{\bsnm{Bouillaud}, \binits{R.}},
\bauthor{\bsnm{Descamps}, \binits{D.}},
\bauthor{\bsnm{Thiele}, \binits{I.}},
\bauthor{\bsnm{Skupin}, \binits{S.}},
\bauthor{\bsnm{Jepsen}, \binits{P.U.}},
\bauthor{\bsnm{Berg\'{e}}, \binits{L.}}:
\batitle{Wavelength scaling of terahertz pulse energies delivered by two-color
  air plasmas}.
\bjtitle{Opt. Lett.}
\bvolume{44}(\bissue{6}),
\bfpage{1488}--\blpage{1491}
(\byear{2019})
\end{barticle}
\endbibitem

\bibitem{Vvedenskii2014}
\begin{barticle}
\bauthor{\bsnm{Vvedenskii}, \binits{N.V.}},
\bauthor{\bsnm{Korytin}, \binits{A.I.}},
\bauthor{\bsnm{Kostin}, \binits{V.A.}},
\bauthor{\bsnm{Murzanev}, \binits{A.A.}},
\bauthor{\bsnm{Silaev}, \binits{A.A.}},
\bauthor{\bsnm{Stepanov}, \binits{A.N.}}:
\batitle{Two-color laser-plasma generation of terahertz radiation using a
  frequency-tunable half harmonic of a femtosecond pulse}.
\bjtitle{Phys. Rev. Lett.}
\bvolume{112},
\bfpage{055004}
(\byear{2014}).
\doiurl{10.1103/PhysRevLett.112.055004}
\end{barticle}
\endbibitem

\bibitem{Kostin2016}
\begin{barticle}
\bauthor{\bsnm{Kostin}, \binits{V.A.}},
\bauthor{\bsnm{Laryushin}, \binits{I.D.}},
\bauthor{\bsnm{Silaev}, \binits{A.A.}},
\bauthor{\bsnm{Vvedenskii}, \binits{N.V.}}:
\batitle{Ionization-induced multiwave mixing: Terahertz generation with
  two-color laser pulses of various frequency ratios}.
\bjtitle{Phys. Rev. Lett.}
\bvolume{117},
\bfpage{035003}
(\byear{2016}).
\doiurl{10.1103/PhysRevLett.117.035003}
\end{barticle}
\endbibitem

\bibitem{Zhang2017}
\begin{barticle}
\bauthor{\bsnm{Zhang}, \binits{L.-L.}},
\bauthor{\bsnm{Wang}, \binits{W.-M.}},
\bauthor{\bsnm{Wu}, \binits{T.}},
\bauthor{\bsnm{Zhang}, \binits{R.}},
\bauthor{\bsnm{Zhang}, \binits{S.-J.}},
\bauthor{\bsnm{Zhang}, \binits{C.-L.}},
\bauthor{\bsnm{Zhang}, \binits{Y.}},
\bauthor{\bsnm{Sheng}, \binits{Z.-M.}},
\bauthor{\bsnm{Zhang}, \binits{X.-C.}}:
\batitle{Observation of terahertz radiation via the two-color laser scheme with
  uncommon frequency ratios}.
\bjtitle{Phys. Rev. Lett.}
\bvolume{119},
\bfpage{235001}
(\byear{2017}).
\doiurl{10.1103/PhysRevLett.119.235001}
\end{barticle}
\endbibitem

\bibitem{Thomson2010}
\begin{barticle}
\bauthor{\bsnm{Thomson}, \binits{M.D.}},
\bauthor{\bsnm{Blank}, \binits{V.}},
\bauthor{\bsnm{Roskos}, \binits{H.G.}}:
\batitle{Terahertz white-light pulses from an air plasma photo-induced by
  incommensurate two-color optical fields}.
\bjtitle{Opt. Express}
\bvolume{18}(\bissue{22}),
\bfpage{23173}--\blpage{23182}
(\byear{2010}).
\doiurl{10.1364/OE.18.023173}
\end{barticle}
\endbibitem

\bibitem{Babushkin2011}
\begin{barticle}
\bauthor{\bsnm{Babushkin}, \binits{I.}},
\bauthor{\bsnm{Skupin}, \binits{S.}},
\bauthor{\bsnm{Husakou}, \binits{A.}},
\bauthor{\bsnm{Köhler}, \binits{C.}},
\bauthor{\bsnm{Cabrera-Granado}, \binits{E.}},
\bauthor{\bsnm{Berg{\'{e}}}, \binits{L.}},
\bauthor{\bsnm{Herrmann}, \binits{J.}}:
\batitle{Tailoring terahertz radiation by controlling tunnel photoionization
  events in gases}.
\bjtitle{New Journal of Physics}
\bvolume{13}(\bissue{12}),
\bfpage{123029}
(\byear{2011}).
\doiurl{10.1088/1367-2630/13/12/123029}
\end{barticle}
\endbibitem

\bibitem{Balciunas2015}
\begin{barticle}
\bauthor{\bsnm{Bal\v{c}i\={u}nas}, \binits{T.}},
\bauthor{\bsnm{Lorenc}, \binits{D.}},
\bauthor{\bsnm{Ivanov}, \binits{M.}},
\bauthor{\bsnm{Smirnova}, \binits{O.}},
\bauthor{\bsnm{Zheltikov}, \binits{A.M.}},
\bauthor{\bsnm{Dietze}, \binits{D.}},
\bauthor{\bsnm{Unterrainer}, \binits{K.}},
\bauthor{\bsnm{Rathje}, \binits{T.}},
\bauthor{\bsnm{Paulus}, \binits{G.G.}},
\bauthor{\bsnm{Baltu\v{s}ka}, \binits{A.}},
\bauthor{\bsnm{Haessler}, \binits{S.}}:
\batitle{Cep-stable tunable thz-emission originating from
  laser-waveform-controlled sub-cycle plasma-electron bursts}.
\bjtitle{Opt. Express}
\bvolume{23}(\bissue{12}),
\bfpage{15278}--\blpage{15289}
(\byear{2015}).
\doiurl{10.1364/OE.23.015278}
\end{barticle}
\endbibitem

\bibitem{Martinez2015}
\begin{barticle}
\bauthor{\bsnm{Mart\'{\i}nez}, \binits{P.G.d.A.}},
\bauthor{\bsnm{Babushkin}, \binits{I.}},
\bauthor{\bsnm{Berg\'e}, \binits{L.}},
\bauthor{\bsnm{Skupin}, \binits{S.}},
\bauthor{\bsnm{Cabrera-Granado}, \binits{E.}},
\bauthor{\bsnm{K\"ohler}, \binits{C.}},
\bauthor{\bsnm{Morgner}, \binits{U.}},
\bauthor{\bsnm{Husakou}, \binits{A.}},
\bauthor{\bsnm{Herrmann}, \binits{J.}}:
\batitle{Boosting terahertz generation in laser-field ionized gases using a
  sawtooth wave shape}.
\bjtitle{Phys. Rev. Lett.}
\bvolume{114},
\bfpage{183901}
(\byear{2015}).
\doiurl{10.1103/PhysRevLett.114.183901}
\end{barticle}
\endbibitem

\bibitem{Matsubara2012}
\begin{barticle}
\bauthor{\bsnm{Matsubara}, \binits{E.}},
\bauthor{\bsnm{Nagai}, \binits{M.}},
\bauthor{\bsnm{Ashida}, \binits{M.}}:
\batitle{Ultrabroadband coherent electric field from far infrared to 200 thz
  using air plasma induced by 10 fs pulses}.
\bjtitle{Applied Physics Letters}
\bvolume{101}(\bissue{1}),
\bfpage{011105}
(\byear{2012}).
\doiurl{10.1063/1.4732524}
\end{barticle}
\endbibitem

\bibitem{Blank2013}
\begin{barticle}
\bauthor{\bsnm{Blank}, \binits{V.}},
\bauthor{\bsnm{Thomson}, \binits{M.D.}},
\bauthor{\bsnm{Roskos}, \binits{H.G.}}:
\batitle{Spatio-spectral characteristics of ultra-broadband {THz} emission from
  two-colour photoexcited gas plasmas and their impact for nonlinear
  spectroscopy}.
\bjtitle{New Journal of Physics}
\bvolume{15}(\bissue{7}),
\bfpage{075023}
(\byear{2013}).
\doiurl{10.1088/1367-2630/15/7/075023}
\end{barticle}
\endbibitem

\bibitem{Xie2006}
\begin{barticle}
\bauthor{\bsnm{Xie}, \binits{X.}},
\bauthor{\bsnm{Dai}, \binits{J.}},
\bauthor{\bsnm{Zhang}, \binits{X.-C.}}:
\batitle{Coherent control of thz wave generation in ambient air}.
\bjtitle{Phys. Rev. Lett.}
\bvolume{96},
\bfpage{075005}
(\byear{2006}).
\doiurl{10.1103/PhysRevLett.96.075005}
\end{barticle}
\endbibitem

\bibitem{Zhang2020}
\begin{barticle}
\bauthor{\bsnm{Zhang}, \binits{K.}},
\bauthor{\bsnm{Zhang}, \binits{Y.}},
\bauthor{\bsnm{Wang}, \binits{X.}},
\bauthor{\bsnm{Yan}, \binits{T.-M.}},
\bauthor{\bsnm{Jiang}, \binits{Y.H.}}:
\batitle{Continuum electron giving birth to terahertz emission}.
\bjtitle{Photon. Res.}
\bvolume{8}(\bissue{6}),
\bfpage{760}--\blpage{767}
(\byear{2020}).
\doiurl{10.1364/PRJ.377408}
\end{barticle}
\endbibitem

\bibitem{Meng2016}
\begin{barticle}
\bauthor{\bsnm{Meng}, \binits{C.}},
\bauthor{\bsnm{Chen}, \binits{W.}},
\bauthor{\bsnm{Wang}, \binits{X.}},
\bauthor{\bsnm{Lü}, \binits{Z.}},
\bauthor{\bsnm{Huang}, \binits{Y.}},
\bauthor{\bsnm{Liu}, \binits{J.}},
\bauthor{\bsnm{Zhang}, \binits{D.}},
\bauthor{\bsnm{Zhao}, \binits{Z.}},
\bauthor{\bsnm{Yuan}, \binits{J.}}:
\batitle{Enhancement of terahertz radiation by using circularly polarized
  two-color laser fields}.
\bjtitle{Applied Physics Letters}
\bvolume{109}(\bissue{13}),
\bfpage{131105}
(\byear{2016}).
\doiurl{10.1063/1.4963883}
\end{barticle}
\endbibitem

\bibitem{Oh2013}
\begin{barticle}
\bauthor{\bsnm{Oh}, \binits{T.I.}},
\bauthor{\bsnm{You}, \binits{Y.S.}},
\bauthor{\bsnm{Jhajj}, \binits{N.}},
\bauthor{\bsnm{Rosenthal}, \binits{E.W.}},
\bauthor{\bsnm{Milchberg}, \binits{H.M.}},
\bauthor{\bsnm{Kim}, \binits{K.Y.}}:
\batitle{Scaling and saturation of high-power terahertz radiation generation in
  two-color laser filamentation}.
\bjtitle{Applied Physics Letters}
\bvolume{102}(\bissue{20}),
\bfpage{201113}
(\byear{2013}).
\doiurl{10.1063/1.4807790}
\end{barticle}
\endbibitem

\bibitem{Kuk2016}
\begin{barticle}
\bauthor{\bsnm{Kuk}, \binits{D.}},
\bauthor{\bsnm{Yoo}, \binits{Y.J.}},
\bauthor{\bsnm{Rosenthal}, \binits{E.W.}},
\bauthor{\bsnm{Jhajj}, \binits{N.}},
\bauthor{\bsnm{Milchberg}, \binits{H.M.}},
\bauthor{\bsnm{Kim}, \binits{K.Y.}}:
\batitle{Generation of scalable terahertz radiation from cylindrically focused
  two-color laser pulses in air}.
\bjtitle{Applied Physics Letters}
\bvolume{108}(\bissue{12}),
\bfpage{121106}
(\byear{2016})
{\href{https://arxiv.org/abs/https://doi.org/10.1063/1.4944843}{{https://doi.org/10.1063/1.4944843}}}.
\doiurl{10.1063/1.4944843}
\end{barticle}
\endbibitem

\bibitem{Zhang2018c}
\begin{barticle}
\bauthor{\bsnm{Zhang}, \binits{Z.}},
\bauthor{\bsnm{Chen}, \binits{Y.}},
\bauthor{\bsnm{Cui}, \binits{S.}},
\bauthor{\bsnm{He}, \binits{F.}},
\bauthor{\bsnm{Chen}, \binits{M.}},
\bauthor{\bsnm{Zhang}, \binits{Z.}},
\bauthor{\bsnm{Yu}, \binits{J.}},
\bauthor{\bsnm{Chen}, \binits{L.}},
\bauthor{\bsnm{Sheng}, \binits{Z.}},
\bauthor{\bsnm{Zhang}, \binits{J.}}:
\batitle{{Manipulation of polarizations for broadband terahertz waves emitted
  from laser plasma filaments}}.
\bjtitle{NATURE PHOTONICS}
\bvolume{12}(\bissue{9}),
\bfpage{554}--\blpage{559}
(\byear{2018}).
\doiurl{10.1038/s41566-018-0238-9}
\end{barticle}
\endbibitem

\bibitem{Sheng2021}
\begin{barticle}
\bauthor{\bsnm{Sheng}, \binits{W.}},
\bauthor{\bsnm{Tang}, \binits{F.}},
\bauthor{\bsnm{Zhang}, \binits{Z.}},
\bauthor{\bsnm{Chen}, \binits{Y.}},
\bauthor{\bsnm{Peng}, \binits{X.-Y.}},
\bauthor{\bsnm{Sheng}, \binits{Z.-M.}}:
\batitle{Spectral control of terahertz radiation from inhomogeneous plasma
  filaments by tailoring two-color laser beams}.
\bjtitle{Opt. Express}
\bvolume{29}(\bissue{6}),
\bfpage{8676}--\blpage{8684}
(\byear{2021}).
\doiurl{10.1364/OE.417515}
\end{barticle}
\endbibitem

\bibitem{Yoo2019}
\begin{barticle}
\bauthor{\bsnm{Yoo}, \binits{Y.-J.}},
\bauthor{\bsnm{Jang}, \binits{D.}},
\bauthor{\bsnm{Kim}, \binits{K.-Y.}}:
\batitle{Highly enhanced terahertz conversion by two-color laser filamentation
  at low gas pressures}.
\bjtitle{Opt. Express}
\bvolume{27}(\bissue{16}),
\bfpage{22663}--\blpage{22673}
(\byear{2019}).
\doiurl{10.1364/OE.27.022663}
\end{barticle}
\endbibitem

\bibitem{He2021}
\begin{barticle}
\bauthor{\bsnm{He}, \binits{T.}},
\bauthor{\bsnm{Zhang}, \binits{Y.}},
\bauthor{\bsnm{Zhao}, \binits{J.J.}},
\bauthor{\bsnm{Wang}, \binits{X.}},
\bauthor{\bsnm{Shen}, \binits{Z.}},
\bauthor{\bsnm{Jin}, \binits{Z.}},
\bauthor{\bsnm{Yan}, \binits{T.-M.}},
\bauthor{\bsnm{Jiang}, \binits{Y.}}:
\batitle{Third-order harmonic generation in a bi-chromatic elliptical laser
  field}.
\bjtitle{Opt. Express}
\bvolume{29}(\bissue{14}),
\bfpage{21936}--\blpage{21946}
(\byear{2021}).
\doiurl{10.1364/OE.427232}
\end{barticle}
\endbibitem

\bibitem{Geissler1999}
\begin{barticle}
\bauthor{\bsnm{Geissler}, \binits{M.}},
\bauthor{\bsnm{Tempea}, \binits{G.}},
\bauthor{\bsnm{Scrinzi}, \binits{A.}},
\bauthor{\bsnm{Schnürer}, \binits{M.}},
\bauthor{\bsnm{Krausz}, \binits{F.}},
\bauthor{\bsnm{Brabec}, \binits{T.}}:
\batitle{Light propagation in field-ionizing media: Extreme nonlinear optics}.
\bjtitle{Phys. Rev. Lett.}
\bvolume{83}(\bissue{15}),
\bfpage{2930}--\blpage{2933}
(\byear{2006}).
\doiurl{10.1103/PhysRevLett.83.2930}
\end{barticle}
\endbibitem

\bibitem{Gaarde08}
\begin{barticle}
\bauthor{\bsnm{Gaarde}, \binits{M.B.}},
\bauthor{\bsnm{Tate}, \binits{J.L.}},
\bauthor{\bsnm{Schafer}, \binits{K.J.}}:
\batitle{Macroscopic aspects of attosecond pulse generation}.
\bjtitle{J. Phys. B}
\bvolume{41}(\bissue{13}),
\bfpage{132001}--\blpage{1320026}
(\byear{2008}).
\doiurl{10.1088/0953-4075/41/13/132001}
\end{barticle}
\endbibitem

\bibitem{Tong05}
\begin{barticle}
\bauthor{\bsnm{Tong}, \binits{X.M.}},
\bauthor{\bsnm{Lin}, \binits{C.D.}}:
\batitle{Empirical formula for static field ionization rates of atoms and
  molecules by lasers in the barrier-suppression regime}.
\bjtitle{J. Phys. B}
\bvolume{38}(\bissue{15}),
\bfpage{2593}--\blpage{2600}
(\byear{2005}).
\doiurl{10.1088/0953-4075/38/15/001}
\end{barticle}
\endbibitem

\bibitem{Kre06}
\begin{barticle}
\bauthor{\bsnm{Kreß}, \binits{M.}},
\bauthor{\bsnm{Löffler}, \binits{T.}},
\bauthor{\bsnm{Thomson}, \binits{M.D.}},
\bauthor{\bsnm{Dörner}, \binits{R.}},
\bauthor{\bsnm{Gimpel}, \binits{H.}},
\bauthor{\bsnm{Zrost}, \binits{K.}},
\bauthor{\bsnm{Ergler}, \binits{T.}},
\bauthor{\bsnm{Moshammer}, \binits{R.}},
\bauthor{\bsnm{Morgner}, \binits{U.}},
\bauthor{\bsnm{Ullrich}, \binits{J.}},
\bauthor{\bsnm{Roskos}, \binits{H.G.}}:
\batitle{Determination of the carrier-envelope phase of few-cycle laser pulses
  with terahertz-emission spectroscopy}.
\bjtitle{Nature Phys.}
\bvolume{2}(\bissue{13}),
\bfpage{327}--\blpage{331}
(\byear{2006}).
\doiurl{10.1038/nphys286}
\end{barticle}
\endbibitem

\bibitem{Constant1999}
\begin{barticle}
\bauthor{\bsnm{Constant}, \binits{E.}},
\bauthor{\bsnm{Garzella}, \binits{D.}},
\bauthor{\bsnm{Breger}, \binits{P.}},
\bauthor{\bsnm{Mével}, \binits{E.}},
\bauthor{\bsnm{Dorrer}, \binits{C.}},
\bauthor{\bsnm{Blanc}, \binits{C.L.}},
\bauthor{\bsnm{Salin}, \binits{F.}},
\bauthor{\bsnm{Agostini}, \binits{P.}}:
\batitle{Optimizing high harmonic generation in absorbing gases: Model and
  experiment}.
\bjtitle{Phys. Rev. Lett.}
\bvolume{82}(\bissue{8}),
\bfpage{1668}--\blpage{1671}
(\byear{1999}).
\doiurl{10.1103/PhysRevLett.82.1668}
\end{barticle}
\endbibitem

\bibitem{Durfee1995}
\begin{barticle}
\bauthor{\bsnm{DurfeeIII}, \binits{C.G.}},
\bauthor{\bsnm{Lynch}, \binits{J.}},
\bauthor{\bsnm{Milchberg}, \binits{H.M.}}:
\batitle{Development of a plasma waveguide for high-intensity laser pulses}.
\bjtitle{Phys. Rev. E}
\bvolume{51}(\bissue{3}),
\bfpage{2368}--\blpage{2388}
(\byear{1995}).
\doiurl{10.1103/PhysRevE.51.2368}
\end{barticle}
\endbibitem

\end{thebibliography}


\end{document}